Fundamentals on Dependence of Volume on Pressure and Temperature

Zi-Kui Liu

Department of Materials Science and Engineering, The Pennsylvania State University,

University Park, Pennsylvania 16802, USA

**Abstract:**

The common wisdom that volume decreases with pressure and increases with temperature is analyzed in terms of Hillert nonequilibrium thermodynamics in the present work. It is shown that the derivative of volume to pressure in a stable system is always negative, i.e., volume decreases with the increase of pressure, when all other natural variables of the system are kept constant. This originates from the stability requirement that the conjugate variables, such as volume and negative pressure, must change in the same direction in a stable system. Consequently, since volume and temperature are not conjugate variables, they do not have to change in the same direction and thus do change in opposite directions in both natural and man-made systems. It is shown that the decrease of volume with the increase of temperature, commonly referred as negative thermal expansion (NTE) in the literature, originates from the statistical competitions of configurations in the system when the volumes of metastable configurations are smaller than that of the ground-state configuration. It is demonstrated that the zentropy theory can concisely explain and accurately predict NTE based on the density functional theory without fitting parameters.

# 1   Introduction

During the "Navrotsky International Symposium" in celebrating her 80th birthday held at the 2023 Sustainable Industrial Processing Summit and Exhibition (SIPS 2023) in Panama, Prof. Navrotsky, the leading expert on the effects of pressure on materials properties [1–3], emphasized that the volume of a substance must decrease with the increase of pressure.  Apparently, some people observed and thus believed the opposite.  Incidentally, there was a publication at the same time on the very same topic concerning the hP8-to-cP4 structural transition in $Ni_3In$ compounds at high temperature and pressure Bertoldi et al. [4].  It discussed the seemingly "anomalous behavior" reported in the literature that the high-temperature and high-pressure polymorph in several $A_3B$ compounds has a larger volume than their normal form [5].  Webb et al. [5] attributed the origin of the "volume anomaly" to significant larger compressibility in the high-temperature and high-pressure polymorph.  On the other hand, by using density-functional-theory (DFT) calculations, Bertoldi et al. [4] pointed out that this "anomalous behavior" was due to the combined effects of temperature and pressure in the experiments rather than the pressure alone and concluded that there was no volume anomaly in those compounds as the volume is larger at high temperature.

However, it is known that the volume in a system does not always increase with temperature.  A well-known example is that the volume of water decreases with the increase of temperature at temperature below 277 K (4°C), and when it freezes into ice, the volume increases further, which is why icebergs float in water.  It is also known from experiments and DFT calculations that the volume of ice decreases with the increase of temperature at temperature below 70 K [6].  This phenomenon is commonly referred as negative thermal expansion (NTE).  In 1897, Guillaume [7] reported the first manmade NTE material, i.e., an $Fe_{65}Ni_{35}$ alloy with zero thermal expansion (ZTE) at room temperature, now commonly called Invar. In 1996, a centennial symposium was held to review the state-of-the-art in the development and understanding of Invar since its discovery [8]. While many theories were developed to describe NTE mechanisms in various materials, aiming for phenomenological interpretation of experimental observations [8–14], the author's group developed a predictive multiscale entropy approach [15] (recently termed zentropy theory [16]) that was able to accurately predict the ZTE and NTE in $Fe_3Pt$ from DFT calculations without fitting parameters [17].

In the present paper, the volume as a function of pressure and temperature is discussed in terms of Hillert nonequilibrium thermodynamics [18,19] and the zentropy theory [16,20,21] by considering stability of an equilibrium system and internal processes inside the system.

## 2    Review of Hillert Nonequilibrium Thermodynamics

A system can typically have three independent exchanges with its surroundings, i.e., heat ($dQ$), work ($dW$), and mass ($dN_i$) for the mole of component $i$, resulting in the internal energy change of the system, $dU$, as follows in terms of the 1st law of thermodynamics

$$dU = dQ + dW + \sum_{i=1}^{c} U_i dN_i \qquad\qquad Eq.\ 1$$

where $U_i$ is the partial internal energy of component $i$. It is noted that Eq. 1 does not contain information whether the system is in its internal equilibrium. Following Hillert [22], one can write the entropy change of the system as follows [21–24]

$$dS = \frac{dQ}{T} + \sum_{i=1}^{c} S_i dN_i + d_{ip}S \qquad\qquad Eq.\ 2$$

where $S_i$ is the partial entropy of component $i$, and $d_{ip}S$ the entropy production due to independent internal processes ($ip$) in the system. Based on the 2$^{nd}$ law of thermodynamics, an independent irreversible internal process must result in a positive entropy production as follows

$$d_{ip}S > 0 \qquad\qquad Eq.\ 3$$

Combining Eq. 1 and Eq. 2 gives the combined law of thermodynamics as follows

$$dU = TdS + dW + \sum_{i=1}^{c} \mu_i dN_i - Td_{ip}S \qquad\qquad Eq.\ 4$$

$$\mu_i = U_i - TS_i = \left(\frac{\partial U}{\partial N_i}\right)_{S, dW=0, d_{ip}S=0, N_{j\neq i}} \qquad\qquad Eq.\ 5$$

where $\mu_i$ is the chemical potential of component $i$ and depends on both partial internal energy and partial entropy of the component. Eq. 4 was named by the present author as Hillert nonequilibrium thermodynamics or Hillert thermodynamics in order to differentiate from Gibbs equilibrium thermodynamics [18,19]. Gibbs equilibrium thermodynamics or Gibbs thermodynamics is with $d_{ip}S = 0$ as follows

$$dU = TdS + dW + \sum_{i=1}^{c} \mu_i dN_i = \sum_{a=1}^{n} Y^a dX^a \qquad\qquad Eq.\ 6$$

where $Y^a$ and $X^a$ represent the pairs of conjugate variables with $Y^a$ for potentials, such as temperature, stress or pressure, electrical and magnetic fields, and chemical potential, and $X^a$ for molar quantities, such as entropy, strain or volume, electrical and magnetic displacements, and moles of components. $X^a$:s are the independent variables of $U$ and are termed as its natural variables [22]. They are related by the following equation

$$Y^a = \left(\frac{\partial U}{\partial X^a}\right)_{X^{b \neq a}} \qquad\qquad Eq.\ 7$$

For simplicity in terms of purpose of the present work, let us consider one independent internal process in the system and write the entropy production by the following 2nd order Taylor expansion [22–24],

$$T d_{ip} S = D d\xi - \frac{1}{2} D_2 (d\xi)^2 \qquad\qquad Eq.\ 8$$

where $D = -\left(\frac{\partial U}{\partial \xi}\right)_{X^a}$ is the driving force for the internal process, $d\xi$ the change of the internal variable that represents the internal process driving by $D$, and $D_2 = \left(\frac{\partial^2 U}{\partial \xi^2}\right)_{X^a}$ is related to the stability and criticality of the system due to the internal process [15]. For systems without any internal processes, i.e., $d_{ip} S = 0$ or $d\xi = 0$, Hillert [22] discussed two types of equilibria: one without driving force for any internal processes, i.e., $D \leq 0$ for a full equilibrium state, and the other with driving force for some internal processes, but those internal processes are not possible due to internal constraints, i.e., $D > 0$ and $d\xi = 0$ so the system is under *freezing-in* conditions with respect to those internal processes.

Through a virtual experiment by moving a molar quantity from one location to another inside the system, denoted by $d\xi = dX_j$, Hillert [22] showed that at equilibrium, the conjugate potential of $X_j$, i.e., $Y_j$, inside the system must have the same value everywhere inside the system, represented by the subscript of the variables, i.e., all the potentials must be homogeneous inside the system. This is shown by the following equation when $dX_j$ is moved from position $'$ with potential $Y_j'$ to position '' with potential $Y_j^{''}$ without any exchange between the system and its surroundings, i.e., $dX^a = 0$ and $dU = 0$,

$$dU = -Td_{ip}S = -Dd\xi = \left(Y_j^{''} - Y_j'\right)dX_j = 0 \qquad Eq.\ 9$$

resulting in $Y_j^{''} = Y_j'$, i.e., potentials are homogeneous everywhere in the system.

The stability of an equilibrium system is determined by the sign of $D_2$. By the same virtual experiment, the following equation can be obtained from Eq. 8

$$dU = -Td_{ip}S = \frac{1}{2}D_2(d\xi)^2 = \frac{1}{2}\frac{\partial^2 U}{\partial \xi^2}(d\xi)^2 \qquad Eq.\ 10$$

As the 2nd law of thermodynamics requires $d_{ip}S \leq 0$ for an equilibrium system, one must have

$$D_2 = \frac{\partial^2 U}{\partial \xi^2} = \frac{\partial^2 U}{\partial X_j^2} = \frac{\partial Y_j}{\partial X_j} \geq 0 \qquad Eq.\ 11$$

i.e., the conjugate pair of variables must change in the direction. At the limit of stability with $\frac{\partial Y_j}{\partial X_j} = 0$, the system becomes unstable with respect to internal fluctuation of $X_j$, resulting in its positive divergence as follows

$$\frac{\partial X_j}{\partial Y_j} = +\infty \qquad Eq.\ 12$$

If the equilibrium is metastable, the system moves to its more stable state without barrier at the limit of stability of $\frac{\partial Y_j}{\partial X_j} = 0$, i.e., the limit of stability represents a saddle point on the energy landscape. If the equilibrium is a globally stable one, the limit of the stability represents the critical point in the system where the system changes from stable to unstable without becoming metastable first, such as the quantum criticality at 0 K [20,25–31]. Since a critical point in an equilibrium system is a zero-dimension geometric feature in a space represented by the $n$ independent variables shown by Eq. 6, all derivatives of free energy up to $(n + 1)^{\text{th}}$ order are zero, and thus defined critical point was termed as invariant critical point (ICP) [32].

## 3    Review of entropy and zentropy

The entropy change of a system is represented by Eq. 2, and its absolute value can be obtained by integration of experimentally measured heat capacity using the $3^{\text{rd}}$ law of thermodynamics stipulating $S = 0$ at $T = 0\ K$ as follows

$$S = \int_0^T \frac{C_P}{T} dT \qquad Eq.\ 13$$

It is commonly referred as Clausius entropy in the literature.

Theoretically, entropy is discussed in terms of statistical mechanics invented by Gibbs [33] based on the foundation established by Clausius, Maxwell, and Boltzmann. Gibbs "imagined a great number of systems of the same nature, but differing in the configurations and velocities which they have at a given instant, and differing not merely infinitesimally, but it may be so as to embrace every conceivable combination of configuration and velocities". The entropy among the configurations is obtained as follows

$$S^{conf} = -k_B \sum_{k=1}^m p^k ln p^k \qquad Eq.\ 14$$

where $m$ is the number of independent configurations, $p^k$ the probability of configuration $k$, and $k_B$ the Boltzmann constant.

It is evident that Eq. 14 does not include the entropy possessed by each configuration in the system. Our zentropy theory stipulates the total entropy of the system as follows [15,16]

$$S = \sum_{k=1}^{m} p^k S^k - k_B \sum_{k=1}^{m} p^k ln p^k = \int_0^T \frac{C_P}{T} dT \qquad Eq.\ 15$$

where $S^k$ is the entropy of configuration $k$. In our initial work [6,17,34,35], Eq. 15 was derived by assuming that the partition function of each configuration should be evaluated from its free energy rather than its internal energy. It was later realized that one should start from Eq. 15 instead [15], which prompted the naming of the approach as zentropy theory [16].

Consequently, Helmholtz energy and statistical mechanics of a canonical ensemble are obtained as follows in terms of the zentropy theory,

$$F = \sum_{k=1}^{m} p^k U^k - TS = \sum_{k=1}^{m} p^k F^k + k_B T \sum_{k=1}^{m} p^k ln p^k \qquad Eq.\ 16$$

$$Z = e^{-\frac{F}{k_B T}} = \sum_{k=1}^{m} Z^k = \sum_{k=1}^{m} e^{-\frac{F^k}{k_B T}} \qquad Eq.\ 17$$

$$p^k = \frac{Z^k}{Z} = \frac{e^{-\frac{F^k}{k_B T}}}{Z} = e^{-\frac{F^k - F}{k_B T}} \qquad Eq.\ 18$$

where $U^k$, $F^k = U^k - TS^k$, and $Z^k$ are the internal energy, Helmholtz energy, and partition function of configuration $k$, respectively, and $Z$ is the partition function of the ensemble or system. It can been seen that Eq. 17 reduces to the standard Gibbs statistical mechanics when $S^k = 0$ so that $F^k = U^k$, i.e., all configurations are pure quantum configurations without additional internal degrees of freedom as discussed in quantum statistical mechanics by Landau and Lifshitz [36]. The quantitative predictions of free energy and phase transition for a number of magnetic materials in terms of zentropy theory are reviewed by the present author [18,20,21], showing remarkable agreement with experimental observations without fitting parameters. The zentropy theory is being extended to predict the free energy and phase transition in $PbTiO_3$ [37,38], melting [39], and superconductors [40], demonstrating very promising results.

## 4    Dependence of volume on pressure

Let us consider hydrostatical work only for the purpose of the present work. Eq. 6 is thus written as follows for an equilibrium system,

$$dU = TdS - PdV + \sum_{i=1}^{c} \mu_i dN_i \qquad Eq.\ 19$$

where $V$ is the volume change of the system, and $-P$ is the negative pressure and the conjugate potential of $V$. From Eq. 11, one obtains

$$\left(\frac{\partial(-P)}{\partial V}\right)_{S,N_i,d_{ip}S=0} = -\left(\frac{\partial P}{\partial V}\right)_{S,N_i,d_{ip}S=0} \geq 0 \qquad Eq.\ 20$$

$$\left(\frac{\partial V}{\partial P}\right)_{S,N_i,d_{ip}S=0} \leq 0 \qquad Eq.\ 21$$

Therefore, for a stable or metastable equilibrium system or a system under freezing-in conditions, its volume must always decrease with the increase of the pressure under constant $S$ and $N_i$. Otherwise, the system is unstable.

Under typical experimental conditions, temperature is controlled from the surroundings rather than entropy. It is thus more convenient to work with Helmholtz energy with the combined law written as

$$dF = -SdT - PdV + \sum_{i=1}^{c} \mu_i dN_i \qquad \text{Eq. 22}$$

with one of the natural variables being potential rather than being all natural variables for the internal energy as shown in Eq. 19. Eq. 21 is then written as

$$\left(\frac{\partial V}{\partial P}\right)_{T, N_i, d_{ip}S=0} \leq 0 \qquad \text{Eq. 23}$$

## 5  Dependence of volume on temperature and Maxwell relation

To study dependence of volume on temperature in an equilibrium system, it is more convenient to use Gibbs energy with temperature and pressure as its natural variables, i.e.,

$$dG = -SdT + VdP + \sum_{i=1}^{c} \mu_i dN_i \qquad \text{Eq. 24}$$

For a closed system with $dN_i = 0$, one has

$$\frac{\partial V}{\partial P} = \frac{\partial^2 G}{\partial P^2} \leq 0 \qquad \text{Eq. 25}$$

$$\frac{\partial S}{\partial T} = -\frac{\partial^2 G}{\partial T^2} \geq 0 \qquad \text{Eq. 26}$$

with the other natural variables kept constant which are omitted for the sake of simplicity of formula. Eq. 25 and Eq. 26 indicate that the surface of G(T,P) is concave as pointed out by Hillert [22].

The Maxwell relation through the 2nd derivative of Gibbs energy can be obtained as follows

$$\frac{\partial V}{\partial T} = \frac{\partial^2 G}{\partial T \partial P} = \frac{\partial S}{\partial (-P)} = -\frac{\partial S}{\partial P} \qquad \text{Eq. 27}$$

As discussed by the present author [19,20], the left side of Eq. 27 is relatively easy to measure experimentally, while the right side is ideal for theoretically investigations and can be predicted by the zentropy theory discussed in Section 3. Therefore, the Maxwell relation represents a perfect integration of experimental and theoretical approaches to the same phenomenon.

Considering that the configurations of the system are composed of ground-state and non-ground-state configurations based on DFT, the volume of the system and its derivative to temperature can be obtained as follows

$$V = \frac{\partial G}{\partial P} = \frac{\partial(-k_B T ln Z)}{\partial P} = -\frac{k_B T}{Z}\frac{\partial Z}{\partial P} = -\frac{k_B T}{Z}\frac{\partial \sum_{k=1}^{m} Z^k}{\partial P} = -\frac{k_B T}{Z}\sum_{k=1}^{m}\frac{\partial Z^k}{\partial P}$$

$$= -k_B T \sum_{k=1}^{m}\frac{Z^k}{Z Z^k}\frac{\partial Z^k}{\partial P} = \sum_{k=1}^{m}\frac{Z^k}{Z}\frac{\partial(-k_B T ln Z^k)}{\partial P} \qquad Eq.\ 28$$

$$= \sum_{k=1}^{m}\frac{Z^k}{Z}\frac{\partial G^k}{\partial P} = \sum_{k=1}^{m} p^k V^k = V^g + \sum_{k=1}^{m} p^k(V^k - V^g)$$

$$\frac{\partial V}{\partial T} = \frac{\partial V^g}{\partial T} + \sum_{k=1}^{m}\left[p^k\frac{\partial(V^k - V^g)}{\partial T} + \frac{\partial p^k}{\partial T}(V^k - V^g)\right] \qquad Eq.\ 29$$

where $V^g$ and $V^k$ are the volumes of the ground-state configuration and configuration $k$, respectively. Both $V^g$ and $V^k$ can be predicted by DFT through quasiharmonic approximation (QHA) as a function of temperature and pressure [41,42], and they increase with the increase of temperature, i.e., $\frac{\partial V^k}{\partial T} > 0$ and $\frac{\partial V^g}{\partial T} > 0$.

It is self-evident from Eq. 28 that the only option for $V$ not to increase with the increase of temperature is with $(V^k - V^g) < 0$, i.e., the volumes of non-ground-state metastable configurations are smaller than that of the ground-state configuration. This can also be seen from Eq. 29 that with $(V^k - V^g) < 0$, $\frac{\partial(V^k - V^g)}{\partial T}$ likely negative, $p^k > 0$, and $\frac{\partial p^k}{\partial T} > 0$ except

$\frac{\partial p^g}{\partial T} < 0$ which does not affect the value of $\frac{\partial V}{\partial T}$ since $(V^g - V^g) = 0$, $\frac{\partial V}{\partial T}$ becomes negative under

the following condition

$$\frac{\partial V^g}{\partial T} < \left| \sum_{k=1}^{m} \left[ p^k \frac{\partial(V^k - V^g)}{\partial T} + \frac{\partial p^k}{\partial T}(V^k - V^g) \right] \right| \qquad \text{Eq. 30}$$

Therefore, the phenomenon of negative thermal expansion is related to statistical competition of configurations with the volumes of non-ground-state configurations being smaller than that of the ground-state configuration. The experimental observed macroscopic soft vibrational modes are subjectively assigned to atomic bonding and do not reflect the microscopic individual local bonding characteristics as recently demonstrated by Wendt et al. [43] through neutron scattering investigations.

Furthermore, as discussed in Section 2 and shown by Eq. 12, both V and S diverge with respect to their respective conjugate variables at the critical point, i.e.,

$$\frac{\partial S}{\partial T} = +\infty \qquad \text{Eq. 31}$$

$$\frac{\partial V}{\partial(-P)} = +\infty \qquad \text{Eq. 32}$$

As the critical point is a zero-dimension feature in a multi-dimensional space [22,32], i.e., the two dimensional T-P space in the present case, V and S also diverge with respect to temperature and pressure, respectively, though their signs are not required to be positive by any fundamental laws or stability criteria. On the other hand, Eq. 28 to Eq. 30 show that their signs are positive if $(V^k - V^g) > 0$ such as Ce [6,34,35] and can be negative if $(V^k - V^g) < 0$ such as Fe3Pt [6,17], i.e., at the critical point, one may have

$$\frac{\partial V}{\partial T} = -\frac{\partial S}{\partial P} = \pm\infty \qquad\qquad Eq.\ 33$$

as predicted by the zentropy theory, shown in Figure 1 in terms of U-V curves at 0 K and T-P and T-V phase diagrams, being positive divergency for Ce with $(V^k - V^g) > 0$ and negative divergency for Fe$_3$Pt with $(V^k - V^g) < 0$, respectively.

Since the critical point separates the 1st- and 2nd-order transitions, the two-phase equilibrium line for the 1st-order transition in the T-P space is with its slope represented by the Clausius–Clapeyron equation, i.e.,

$$\frac{\partial P}{\partial T} = \frac{\Delta S}{\Delta V} \qquad\qquad Eq.\ 34$$

with $\Delta S$ and $\Delta V$ being the entropy and volume differences of the two phases or states with the different probabilities for each configuration. A positive slope indicates that $\Delta S$ and $\Delta V$ have the same sign so the high temperature state with high entropy has larger volume than the low temperature state with low entropy (see Eq. 26), resulting in positive sign for Eq. 33 at the critical point. This is shown in Figure 1 (b) for Ce. On the other hand, a negative slope indicates that the high temperature state has smaller volume, thus negative sign for Eq. 33 at the critical point. Consequently, one can determine whether a system may possess negative thermal expansion based on the slopy of a two-phase equilibrium as surveyed in our previous work [44]. This is shown in Figure 1 (e) for Fe3Pt, noting that the pressure is plotted in the descending order from left to right. This approach can be useful to the investigation of phase relations under high temperature and high pressure which are among core activities of Prof. Navrotsky [1–3].

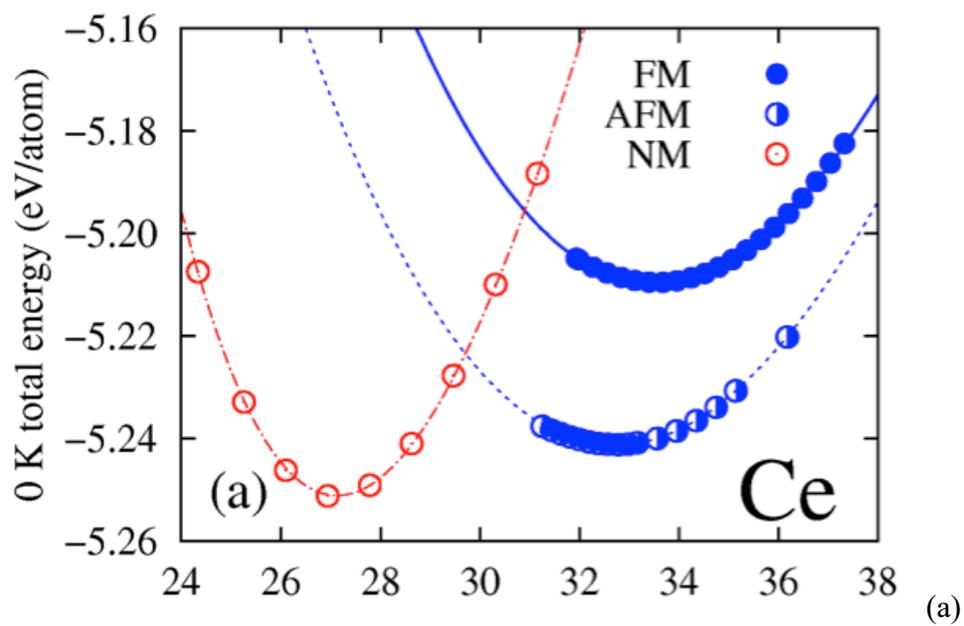

(a)

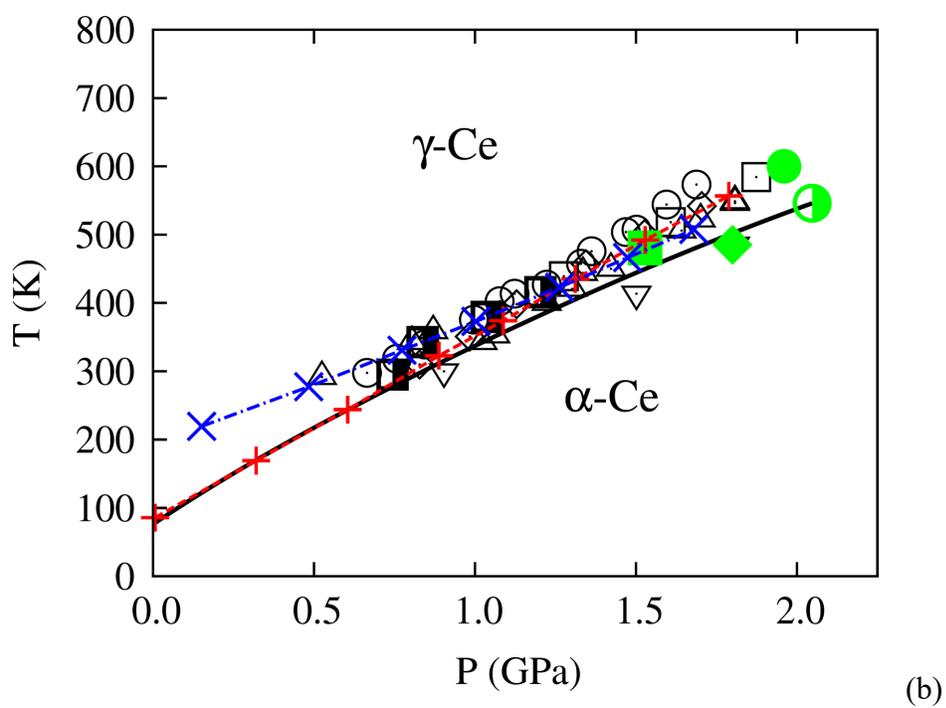

(b)

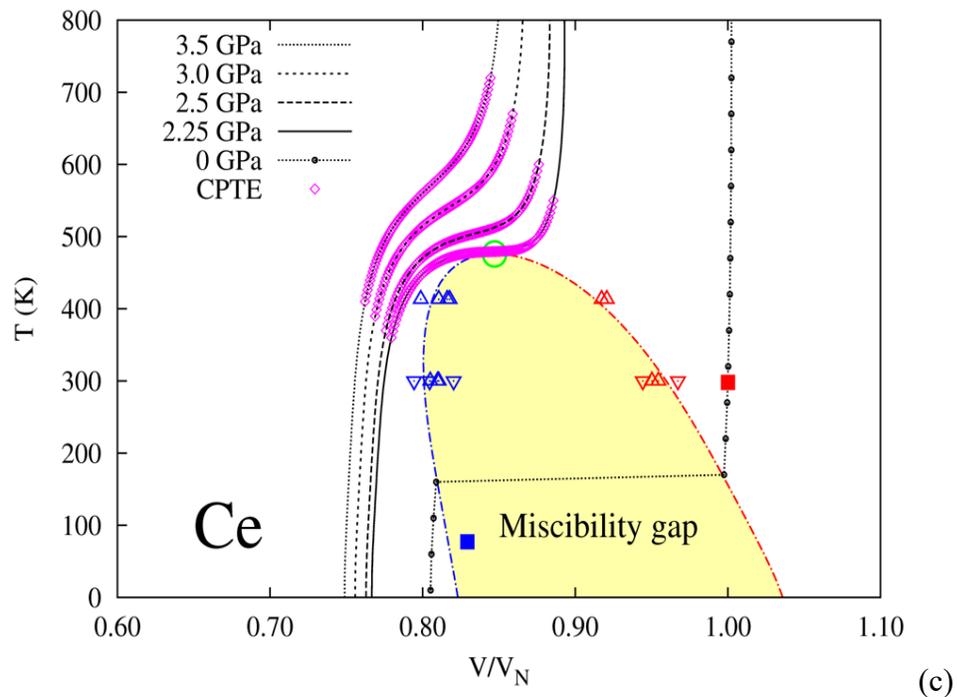

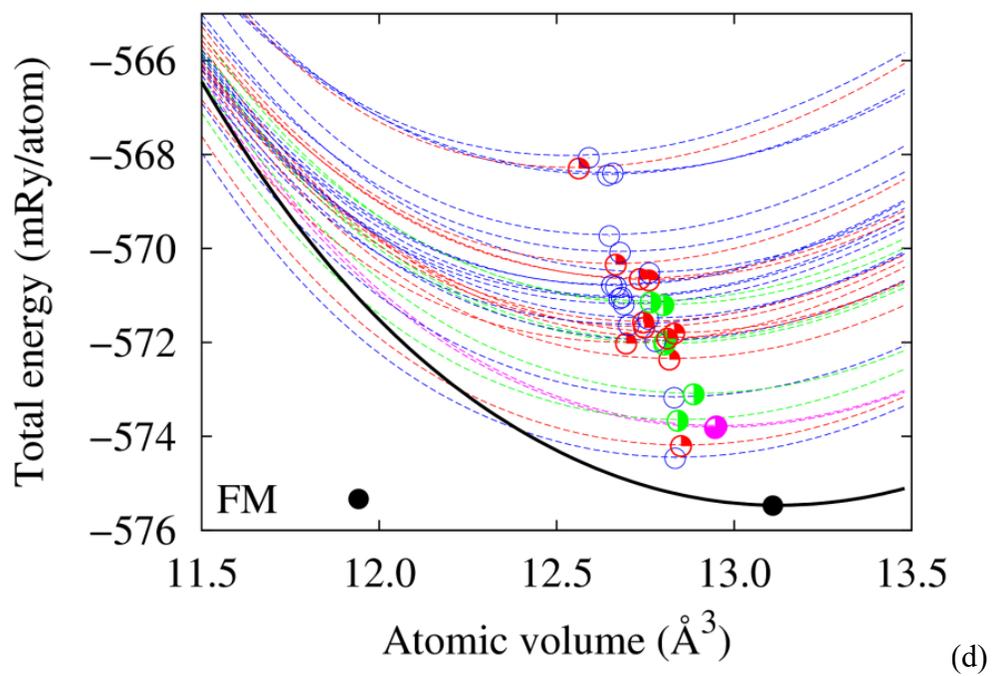

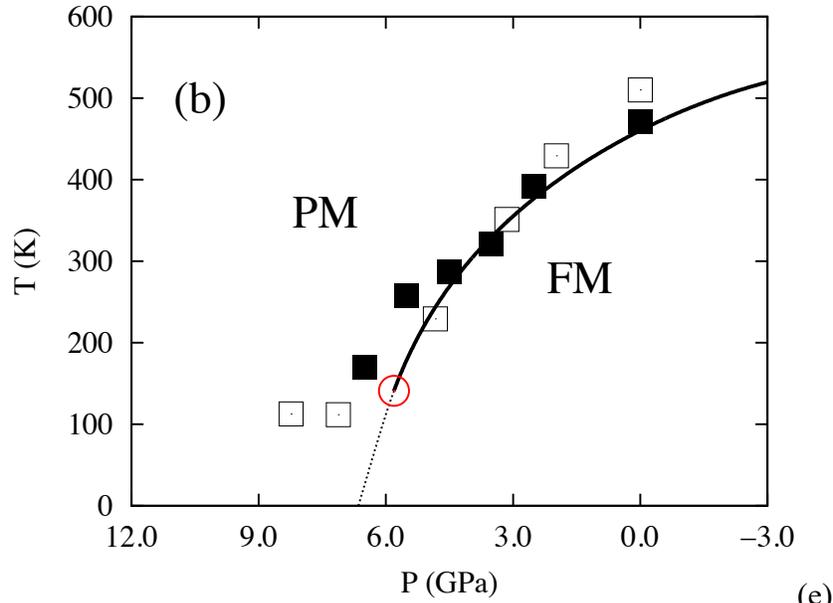

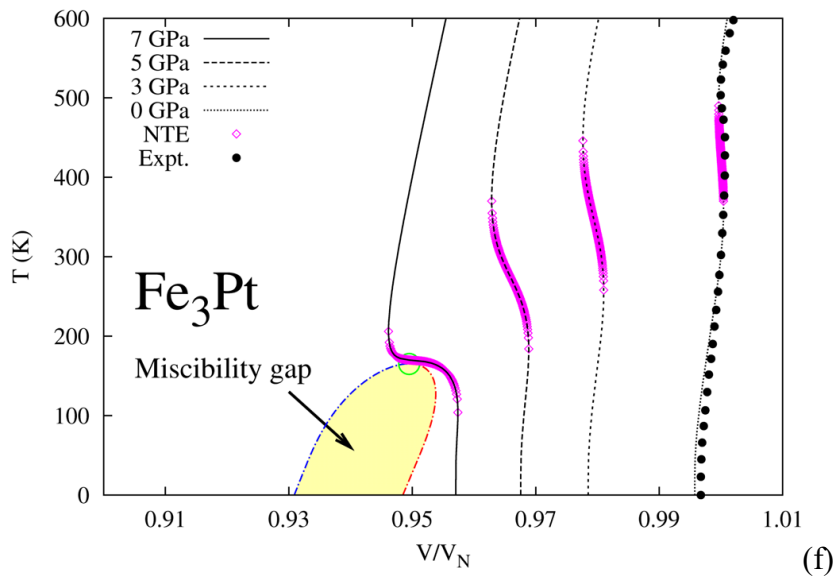

Figure 1: *Predicted U-V at 0K and T-P and T-V phase diagrams of Ce* [16,35] *in (a), (b), and (c), and Fe₃Pt* [16,17] *in (d), (e), and (f) in terms of the zentropy approach. The critical point id denoted by the half-filled circle in (b) and the open red circle in (e) and the green circles in (c) and (f), respectively, and all other symbols are experimental data. The isobaric volume curves are also plotted in (d) and (f) as a function of temperature with the T-P ranges for volume anomaly*

*marked by purple open diamond symbols. Reproduced with permissions from J. Phys. Condens.*

*Matter 21, 326003 (2009) for (a) and (b)* [35]*, Copyright 2009 IOP Publishing, Ltd; from Liu, Z.*

*K., Wang, Y. & Shang, S.-L. Zentropy Theory for Positive and Negative Thermal Expansion, J.*

*Phase Equilibria Diffus. 43, 598–605 (2022) licensed under a Creative Commons license for (c)*

*and (f)* [16]*; and from Philos. Mag. Lett. 90, 851–859 (2010) for (d) and (e)* [17]*, Copyright 2010*

*Taylor & Francis.*

## 6    Summary

In the present paper, the dependence of volume on pressure and temperature is analyzed in the framework of Hillert nonequilibrium thermodynamics through internal processes.  Through virtual experiments, it is shown that the conjugate variables in the combined law of thermodynamics change in the same direction in a stable system, including volume and its conjugate potential, i.e., negative pressure.  Consequently, volume must always decrease with the increase of pressure when other independent variables are kept constant.  On the other hand, since volume and temperature are not conjugate variables, they can change in opposite directions, resulting in NTE in many systems.  It is demonstrated that NTE can be qualitatively forecasted from potential phase diagrams and quantitatively predicted by zentropy theory with inputs solely from DFT without fitting parameters.

## 7    Acknowledgements

The author would like to thank Alexandra Navrotsky for many inspiring lectures and stimulating discussions over the years and admire her significant contributions to the scientific community in general and thermodynamics in particular. The present review article covers research outcomes supported by multiple funding agencies over many years with the most recent ones including U.S. Department of Energy (DOE) Grant No. DE-SC0023185, DE-AR0001435, DE-NE0008945, and DE-NE0009288, and U.S. National Science Foundation (NSF) Grant No. NSF-2229690.